# Sound Absorption by Acoustic Microlattice with Optimized Pore Configuration

Xiaobing Cai and Jun Yang[a]

Department of Mechanical and Materials Engineering, The University of Western Ontario,

London, Ontario, Canada N6A 5B9.

caixiaobing11@gmail.com, jyang@eng.uwo.ca

Gengkai Hu

School of Aerospace Engineering, Beijing Institute of Technology, Beijing 100081, People's

Republic of China.

hugeng@bit.edu.cn

Tianjian Lu

MOE Key Laboratory for Multifunctional Materials and Structures, Xi'an Jiaotong University,

Xi'an, Shanxi, P.R. China, 710049.

State Key Laboratory for Strength and Vibration of Mechanical Structures, Xi'an Jiaotong

University, Xi'an, Shanxi, P.R. China, 710049.

tjlu@xjtu.edu.cn

Running title: Optimal sound absorption by microlattice materials

[a] Author to whom correspondence should be addressed.

**Abstract**: Sound absorption or dissipation principally involves joint interactions between sound waves, material morphology and the air medium. How these elements work most efficiently for sound absorption remains elusive to date. In this paper, we suggest a fundamental relation concisely cross-linking the three elements, which reveals that optimal sound absorption efficiency occurs when the pore size of the material is twice the thickness of the viscous boundary layer of the acoustic air medium. The study is validated by microlattice materials comprising of well-controlled regular structures that absorb sound in a tunable manner. Optimized material morphology in terms of pore size and porosity is determined to provide a robust guidance for optimizing sound absorbing materials.

## 1. Introduction

Sound absorbing materials (SAM) are widely employed for the elimination of noises in many applications, including buildings, vehicles, engines, electric devices and medical instruments. Over the last decades, significant progress has been made in the development of novel SAMs, including perforated panels[1], nano-porous/fibrous materials[2], biomimetic materials[3] and piezoelectric materials[4] and etc. However, the absorption efficiency of these materials for low frequency sound waves is fundamentally limited, so that bulky structures are required to absorb low frequency noises. Therefore, there is a need to optimize the structures of SAMs to overcome the limitations of current sound absorption technologies to make them light and compact.

The mechanisms of sound dissipation in SAMs involve joint interactions between characteristics of the sound wave (e.g. frequency), material morphology (e.g. pore size and porosity), and characteristics of the air medium (e.g. viscosity), along with effects from many other parameters, such as thermal conductivity and elasticity[1, 5-6]. However, the implicit and complicated nature of these interactions makes it challenging to optimize the interior structure of SAMs for more efficient sound absorption. For example, changing the pore size has counteracting effects on the efficiency of sound absorption. On one side, smaller pores would result in enhanced airflow friction and thus increased sound dissipation. On the other hand, smaller pores would also narrow down the airflow passage and thus increase airflow resistance, which favors reflection rather than absorption. The counteracting effects of pore size on sound absorption suggest the existence of an optimal pore size for maximal sound absorption. However, due to limited structural control during SAM synthesis and/or lack of accuracy in micro-fabrication, few prior works studied sound dissipation in SAMs with well-organized pores. Instead, many previous works focused on characterization of the viscous-thermal effects of SAMs with irregular porous networks[7-13], leaving the optimal structural condition for sound absorption unsolved. In this work, we studied sound absorption in microlattice materials[14-19] with well-defined micro-structures to theoretically and experimentally determine the optimal condition. According to our analysis, the microlattice is most absorptive when its pore size is twice the thickness of the viscous boundary layer.

## 2. Microlattice metamaterial of controlled microstructures

The proposed microlattice material is comprised of multiple layers of perforated membranes, spaced apart from each other with air gaps and interconnected by micro-sized rods. As shown in Fig. 1, each monolayer of the membranes is formed by micro-ridges and wires, defining uniform sub-millimeter square pores. The rods and ridges, thicker and wider than the wires, act as solid and sturdy frame of the microlattice. As circular samples are prepared for facilitating the absorption coefficient measurement, each sample has a peripheral wrapper layer. The morphological features of the microlattice can be precisely controlled by tuning the following parameters: pore width $a$, wire width $w$, ridge width $s$, square rod width $W$, layer thickness $t$ and layer spacing $h$.

For a perforated layer, the general form of its acoustic impedance can be expressed as[1]

$$z = \kappa + j\chi, \tag{1}$$

where $\kappa = (32\mu t)k_r/(\sigma\rho_0 c_0 d^2)$, $\chi = (\omega t)k_m/(\sigma c_0)$, $j^2 = -1$, $t$ is the layer thickness, and $c_0$ is the sound speed in air, $\mu$ is dynamic viscosity of air, the while $k_r$ and $k_m$ are two variables dependent on pore diameter $d$. The first term $\kappa$ denotes the part of energy dissipation, which can lead to a maximum sound absorption coefficient of $\alpha = 4\kappa/(1+\kappa)^2$. In particular, $\kappa = 1$ results in full absorption, i.e. $\alpha = 1$. This condition yields

$$\sigma d^2 = (32\mu t)k_r/(\rho_0 c_0). \tag{2}$$

The exact expression of $k_r$ is very complicated, but in our case, its value is within the range of 1.0 ~ 1.2.[1] Equation (2) can be further written as:

$$r = \hat{k}\sqrt{2\mu/(\omega\rho_0)}, \qquad (3)$$

with $\hat{k} = \sqrt{8\pi k_r tf /(\sigma c_0)}$, where $f$ is the sound frequency. In out study, it is found that the value of $\hat{k}$ is approximately 1.0 ($\hat{k} = 1.0 \pm 0.07$) if the maximum sound absorption is achieved. This will be discussed in detail in the following. Therefore, it can be reasonably concluded that the optimal pore radius is:

$$r_{\text{Opt}} \approx \sqrt{2\mu/(\omega\rho_0)}, \qquad (4)$$

equaling to the thickness of viscous boundary layer $t_v = \sqrt{2\mu/(\omega\rho_0)}$ [7, 20-21]. Equation (4) may be derived from other models, such as Johnson-Allard model. In a concise form, Eq. (4) reveals a governing relation among material structure (pore size $r$), air medium properties (density $\rho_0$ and dynamic viscosity $\mu$), and the fundamental parameter of sound wave itself ($\omega$), to enable the material mostly absorptive. To our best knowledge, Eq. (4) has never been explicitly given in previously studies, although the influence of viscous boundary layer on sound absorption is widely mentioned[7, 20-21]. Previously studies usually focused on porous material with irregular and non-uniform pores, few researches paid attention to an optimal pore size. In addition, it was a great challenge to find a meaningful value of $\hat{k} = \sqrt{8\pi k_r tf /(\sigma c_0)}$ when internal structure of the material was irregular and random.

The significance of viscous boundary layer on optimal sound absorption may be explained from the aspect of permeability and dissipativity of the pores. As illustrated in Fig. 2, acoustic air medium would become less resistant and less dissipative when it is far away from solid-air interface[22]. The concept of viscous boundary layer *per se*, denotes that viscosity (so as to resistance and dissipativity) may be neglected beyond a certain distance from the solid-air interface[21-22]. Thus inside a pore, starting from its perimeter and radially inwardly towards its center point, the resistance and dissipativity gradually decrease. If $r > \bar{t}_v$, the center region of the pore is beyond the realm of the viscous boundary layer, and hence this center space is somewhat wasteful in the sense of dissipation. By contrary, if $r < \bar{t}_v$, the viscous boundary layers are overlapped in the center region and the resistance is magnified, making this region too resistant to penetrate, which is disadvantageous for sound absorption. Consequently, the ideal situation must be $r = \bar{t}_v$, a case where the two viscous boundary layers precisely fill in the pore, neither overlapping nor leaving any vacancy.

Now, to determine the value of $\hat{k}$, the distributions of sound absorption coefficient $\alpha$ with respect to pore size and porosity, at each of the twelve 1/3 octave frequencies between 200 and 2500 Hz, are obtained and shown as contour plot in Fig. 3. The calculation of sound absorptivity of the microlattice material is carried out by using an integrated transfer matrix method[19]. First, the complex acoustic wave number $k_n$ and impedance $z_n$ of the $n$-th monolayer of the membranes are obtained according to the simulated acoustic response by COMSOL Acoustics. Then a transfer matrix $[t_n]$ of the $n$-th monolayer is developed to establish the relationship of the acoustic pressure and velocity fields between the front and back sides of this layer: $[p_\sigma \, u_\sigma] = [t_n][p_\delta \, u_\delta]$. Lastly, an overall transfer matrix for the whole sample is established as $[T] = [t_1][t_2]\cdots[t_n]\cdots[t_N]$, with $N$ being the total number of layers. As a result, the sound absorption coefficient can be calculated by using $\alpha = 1 - |R|^2$, where $R$ is the reflectance, $r = (T_{11} - z_0 T_{21})/(T_{11} + z_0 T_{21})$, and $z_0$ is the characteristic impedance of air. Sound absorption coefficients over a certain frequency band at various combination of pore sizes and porosities are calculated. The size range of square pores investigated is between 0.05 mm and 0.5 mm (the range of equivalent diameter of circular pore $d=2a/\sqrt{\pi}$ : 0.056-0.56 mm), and the range of porosity is between 0.1 and 0.8. Thickness of each layer is chosen as 50 μm, while the spacing $h$ separating two adjacent layers is selected as 450 μm. Correspondingly, with 100 layers of membranes, the overall thickness of each sample is 50 mm.

By using the values of $f$ and $\sigma$ that correspond to the maximum sound absorption regions circled in Fig. 3, and using the expression $\hat{k} = \sqrt{8\pi k_r t f /(\sigma c_0)}$, it can be found that the value of $\hat{k}$ always closes to 1.0. With the sample thickness changes, only small variation occurs to the value of $\hat{k}$. It is further found that $\hat{k} = 1.0 \pm 0.07$ works for all thickness between 30mm to 100mm. Figure 4 shows the values of $\hat{k}$ as a function of frequency for various sample thickness. Some values of $\hat{k}$ are not provided because of either extremely high computational cost, or, for thick samples, no optimal pore size and porosity at high frequencies.

Figure 3 and Eq. (4) only reveals optimal pore size at specific frequency points, universal optimal pore radius over a frequency band should also be determined. By further rewriting Eq. (4) as: $r_{Opt}^2 = \nu/(\pi f)$, with $\nu = \mu/\rho_0$ being the momentum diffusivity. It reduces to

$$S_{Opt} = \nu T, \tag{5}$$

where $S_{Opt} = \pi r_{Opt}^2$ is the optimal pore area and $T = 1/f$ is the cycle of the sound wave. Equation (5) can be interpreted as: the optimal pore should be configured to allow acoustic airflow momentum to diffuse over its entire area within a cycle. Again, to our best knowledge, this equation has never been reported before. Since $\nu$ is typically constant, the cycle time $T$ becomes the only parameter determining the diffusion effect of acoustic airflow. So for a noise comprising a series of frequencies $f_i$ or cycles $T_i = 1/f_i$, each corresponding to an area $S_{Opt,i} = \nu T_i$, the total area needed for all momentum diffusion would be obtained as $S = \sum_I vT_i = \sum_I S_{\text{Opt},i}$, with $I$ the total number of constituting frequencies. On average, the amount of momentum diffusion over all the constituting frequencies is $\bar{S} = S/I = \sum_I S_{\text{Opt},i}/I$. If this amount of momentum diffusion is considered to be achieved at a single pore size, it would be $\bar{r} = \sqrt{\bar{S}/\pi}$.

This finding can be validated from the overall rating of the microlattice material. One widely used rating of sound absorption is the Standard Sound Absorption Average ($\alpha_{\text{SAA}}$)[23], which incorporates and averages sound coefficients of the twelve 1/3 octave frequencies. Figure 5 shows the SAA mapping of the microlattice over pore size and porosity. A region of remarkable maximum $\alpha_{\text{SAA}}$ (0.72) can be noted at a pore size (diameter) around 180 μm ($r_{\text{Opt}} = 90\,\mu\text{m}$) and a porosity above 0.56. The calculated optimal radius using $\bar{r} = \sqrt{\sum_I r_i^2/I}$ is about 95 μm, which is in good agreement with that derived from Fig. 5.

Further from Fig. 5, a thin line indicating universal optimal combination of pore size and porosity is shown. An expression of this curve is found to be $\sigma d^2 = \breve{c}_2 k_r$, according to curve fitting, with $\breve{c}_2$ a constant simply determined by using initial values: $\breve{c}_2 = 0.56*0.18^2$. This expression can also be obtained from Eq. (2), by defining $\breve{c}_2 = (32\mu t)/(\rho_0 c_0)$ and by using $k_r = 1.0$, as discussed above.

## 3. Verification and comparison

Based on the analysis above, the ideal diameter of the pores of the microlattice materials for experimental study should be between $d = 180$ μm and $d = 190$ μm. The equivalent pore width of square pores is between $a = 160$ μm and $a = 168$ μm. In order to facilitate fabrication without losing importance, the square pore width is selected as an integer multiple times of 39 μm, which is the XY in-plane resolution of the used Asiga 3D printer. Hence, the pore width of $a = 195$ μm (i.e. $d = 224$ μm) was chosen. As shown in Fig. 5, when the porosity is chosen to be 0.56, a pore diameter varying from 180 μm to 224 μm will not result in a distinct change to SAA. Another microlattice sample having a larger pore width $a = 468$ μm (i.e. $d = 528$ μm) and $\sigma = 0.56$ was also fabricated for comparison. The cross-sectional planes of each microlattice sample were directly constructed from a digital matrix, with each element of the microlattice encoded into pixels. During printing, the exposure time for the photosensitive polymer and other machine parameters were carefully adjusted. Upon completion, the printed sample was transferred from the printer's building platform into ethyl alcohol for ultrasonic treatment for 30 seconds to remove adhering uncured resin.

As shown in Fig. 1d and Fig. 6, the samples have uniformly sized pores that are formed by micro wires and ridges. Pores of uniform width $a = 190$ μm (slightly smaller than the designed value of 195 μm) and wires of 45 μm were fabricated, (Fig. 6c). Taking the solid areas occupied by rods and peripheral ring into account, the actual in-plane porosity for each layer is 56%. Scanning electron microscope (SEM) images offer details of the layers and rods (Fig. 6d), showing an excellent structural control in microfabrication as designed. Seamless bonds between rods and ridges can be observed, which constitute a solid and stable framework of the sample. Taking the space between two adjacent layers into account, the volumetric porosity of the fabricated sample is estimated as high as 94%, resulting in extremely small interlayer airflow resistance, and hence good impedance match between air medium and the sample.

To experimentally characterize the sound absorbing efficiency of these 3D-printed microlattice materials, bulk samples were prepared as circular disks (Fig. 6a) with a diameter of 9.84 mm and a height of 5 mm (i.e., 10 layers of membranes). A stack of 10 disks (total thickness $H = 50$ mm) was mounted into a sound impedance tube having the same inner diameter ($D = 9.84$ mm), backed by a rigid steel plate of thickness 5 mm (Fig. 6b). This impedance tube constitutes a loading portion of a standard sound absorption coefficient measuring device[24], which retrieves a material's sound absorption coefficient from acoustic pressure recorded at two relative locations in front of the sample.

The measured sound absorption properties of microlattice samples are shown in Fig. 7a. The measured sound absorption coefficients of microlattice samples (dashed lines) agree well with numerical results predicted using the integrated transfer matrix method (straight line with square or circular symbols). The microlattice material samples with optimal pore configuration show superior performance in sound absorption. The performance of the present microlattice materials with optimal structures was compared with that of most of traditional sound absorbing materials, shown in Fig. 7b. The optimized microlattice ($a = 190$ μm, $\sigma = 0.56$) remarkably surpasses the performance of the glass fiber ($\rho \approx 32$ mg/cm$^3$ [8]) (dashed line), at all frequencies, despite glass fiber being widely recognized as a superb sound absorbing material. Over all frequencies, the absorption coefficients of popularly used sound absorbing foams are generally 0.3-0.4 lower than that of the microlattice materials. The remarkable enhancement in absorptivity of microlattice with optimal microstructures opens new avenue for design and manufacturing of next generation sound absorbing materials.

## 4. Conclusion

Through theoretical analysis, numerical and experimental investigation on sound absorption by microlattice material with well defined microstructures, it is found that the optimal pore size for maximum sound absorption is twice the viscous boundary layer thickness. This condition was yet not explicitly reported before, as to our best knowledge. In addition, optimal combinations of pore size and porosity for maximal sound absorption over the entire frequency band were also determined for practical applications. The findings in this work offer a new design rule for making high performance sound absorbing materials.

**Figure caption**

Fig. 1. (Color online) The structure of microlattice material comprising multi-layers of membranes, connected and supported by micro rods and ridges, (a) 3D digital model of circular disk-shaped sample containing microlattice structure, (b) connection details of micro rods, ridges and wires, (c) forming of square micro-pores by ridges and wires, (d) an image of a disk-shaped sample.

Fig. 2. (Color online) Schematic illustration of the penetrability and dissipativity of a pore with radius smaller, equal or larger than the thickness of viscous boundary layer.

Fig. 3. (Color online) The mapping of sound absorption coefficient of 50 mm thick microlattice with respect to different pore size and porosity at each of 12 octave frequencies.

Fig. 4 The values of $\hat{k}$ in Eq.(3) are approximately around 1.0 ($\pm$ 0.07) when the maximum sound absorption is achieved for various sample thicknesses.

Fig. 5 The distribution sound absorption average (SAA) over 12 octave frequencies of microlattice of 50 mm thickness.

Fig. 6. (Color online) The internal structure of the microlattice under microscope and SEM, (a) the perspective view of the 3d printed samples, (b) the sound absorption measuring device with an upgraded loading portion, (c) size-controlled and uniform pores, (d) solid frameworks of the microlattice.

Fig. 7. (Color online) Sound absorbing properties of microlattice materials. (a) Numerical and experimental results of sound absorbing coefficient agree with each other for microlattice samples with width $a$ = 190 μm (i.e. d ≈ 210 μm) and $a$ =468 μm (i.e. $d$ ≈ 500 μm). (b) Comparison of absorbing coefficients between the microlattice materials and other SAMs (sound absorbing materials).

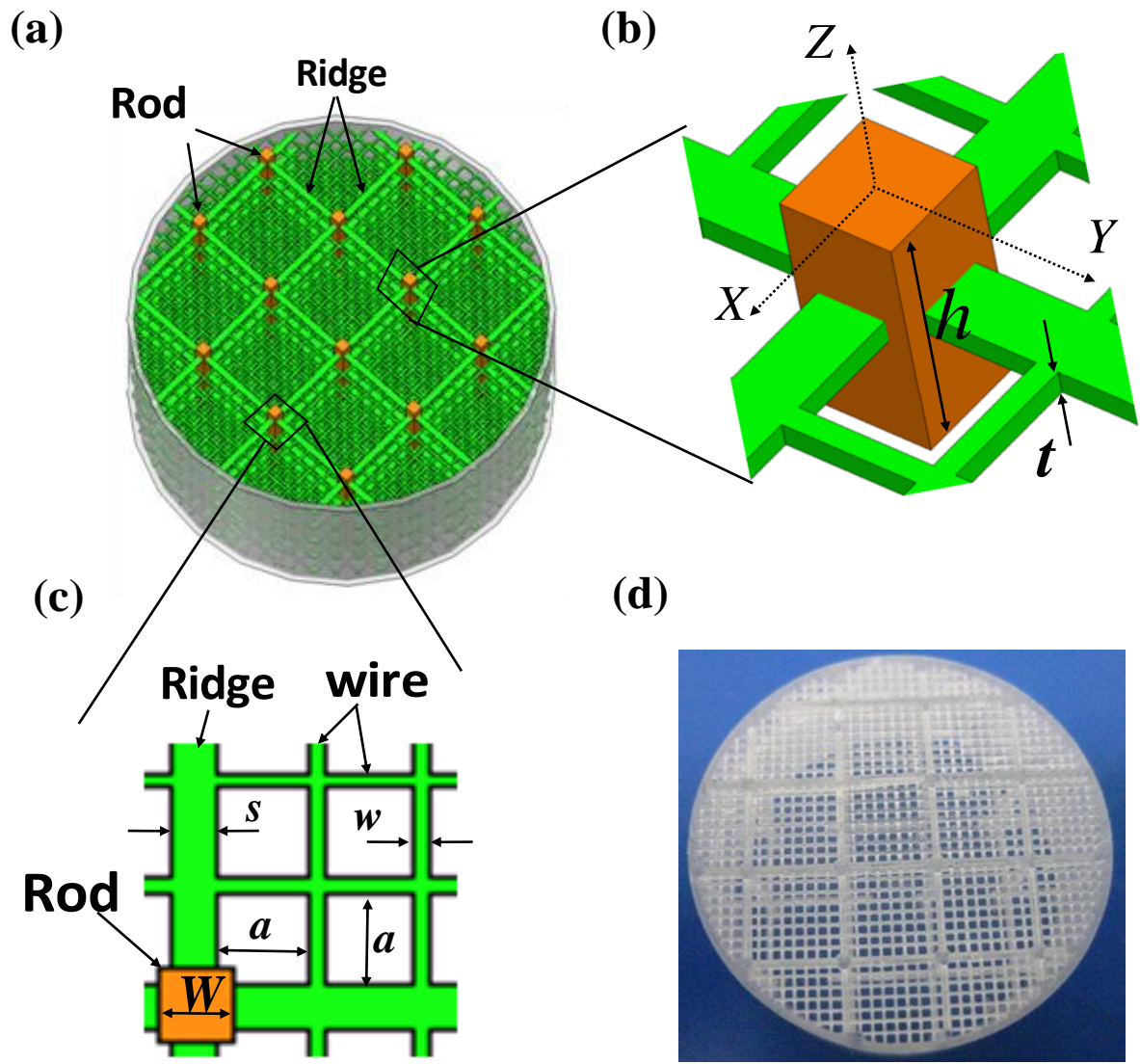

FIG.1

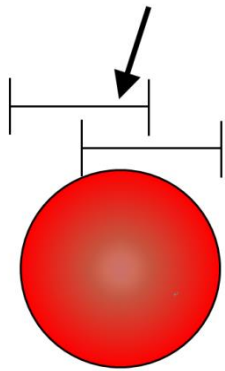 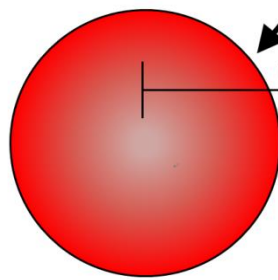 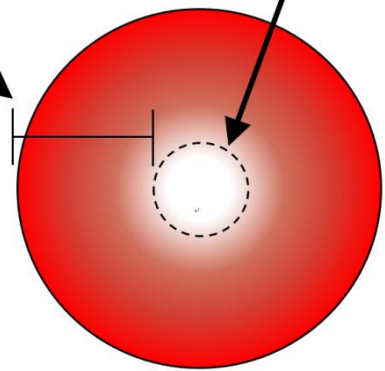

FIG.2

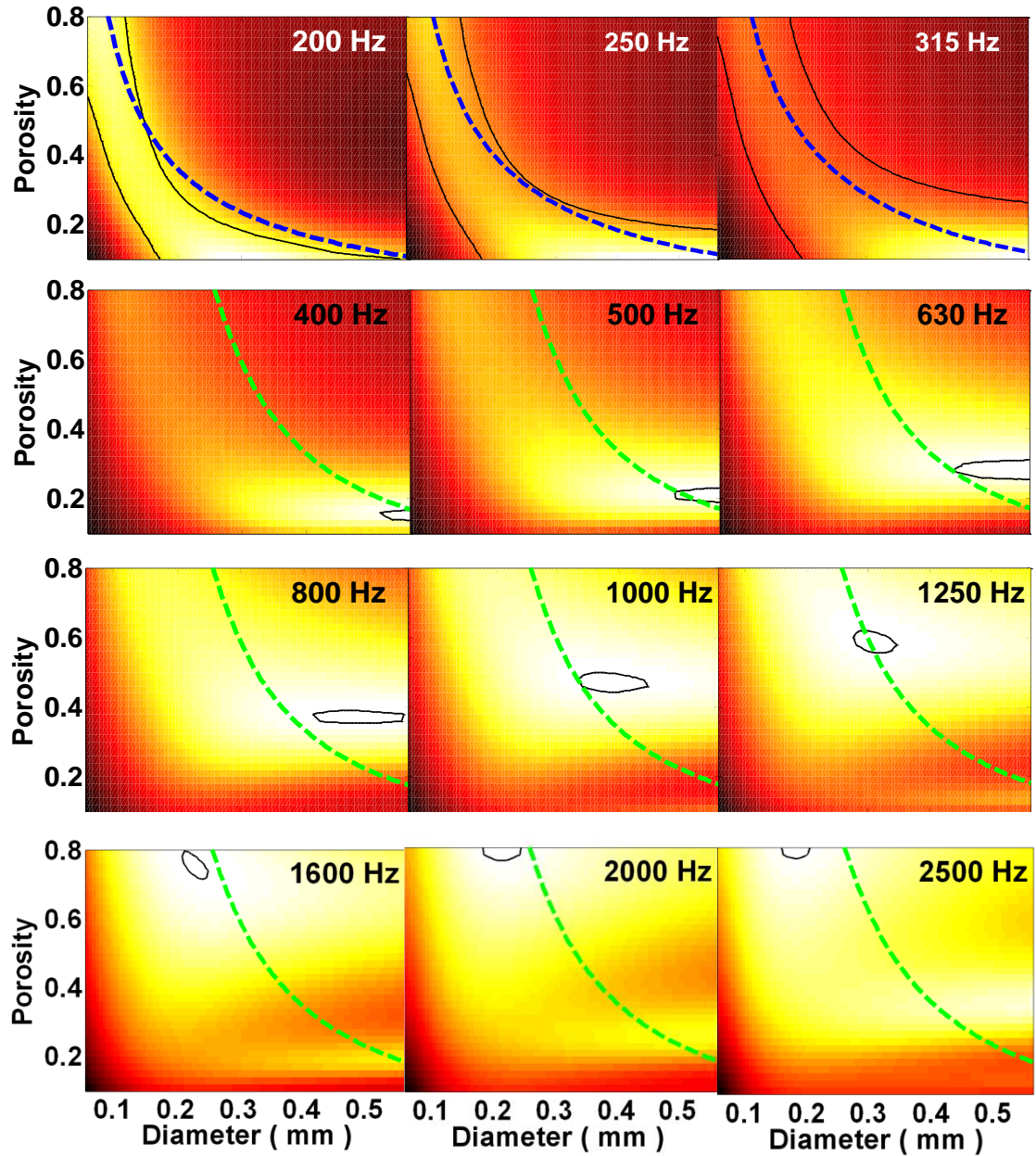

FIG.3

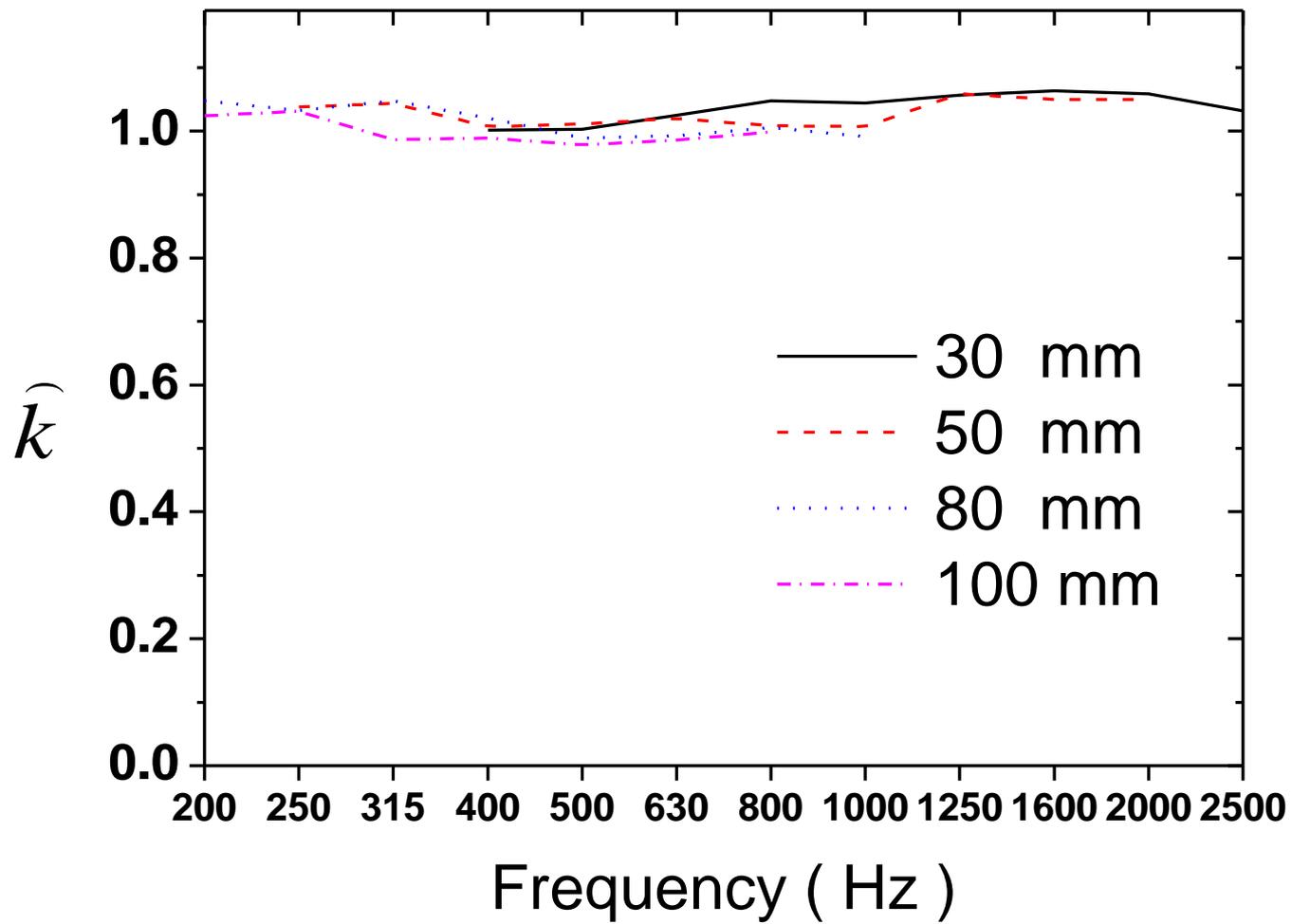

FIG.4

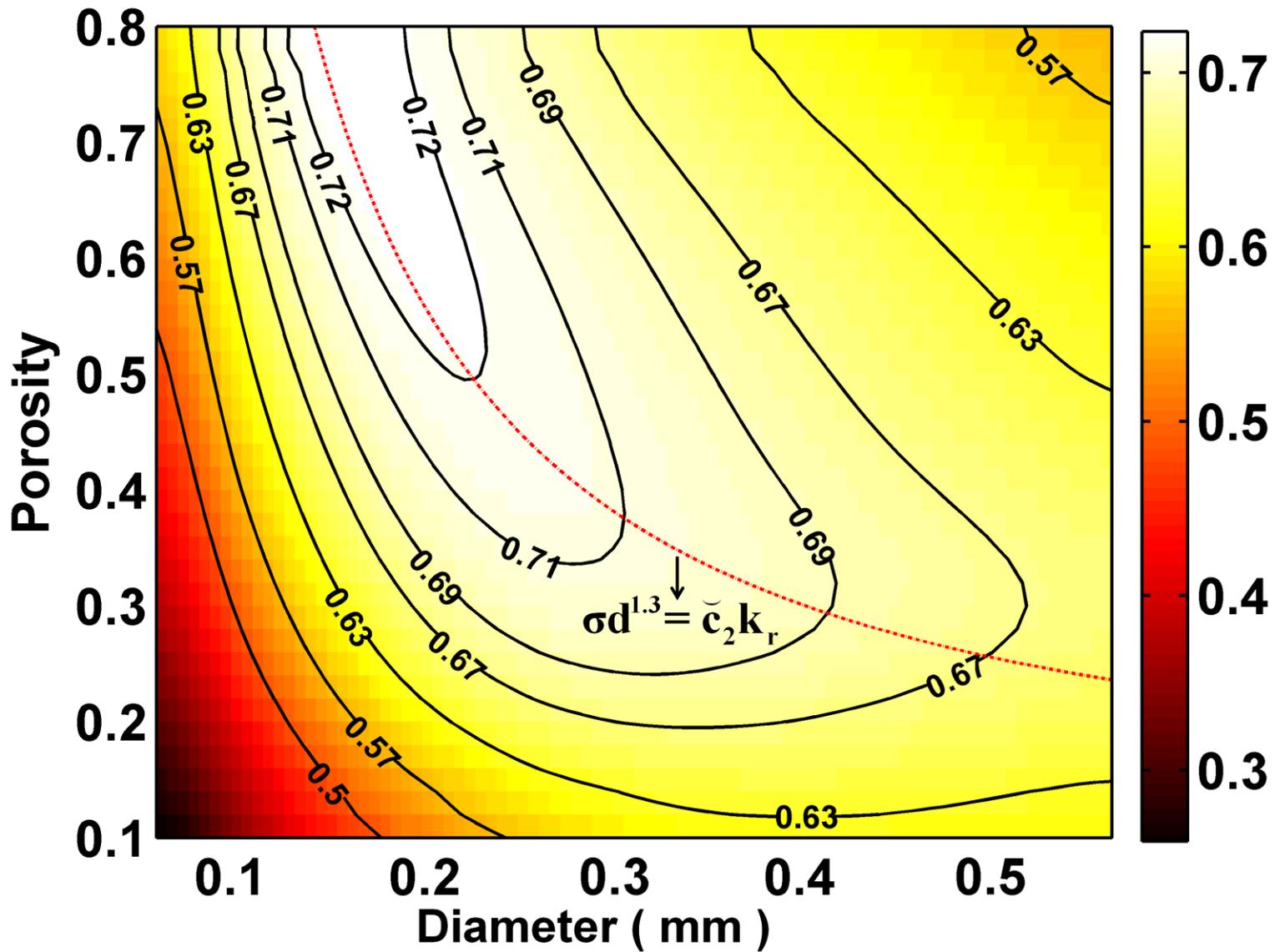

FIG.5

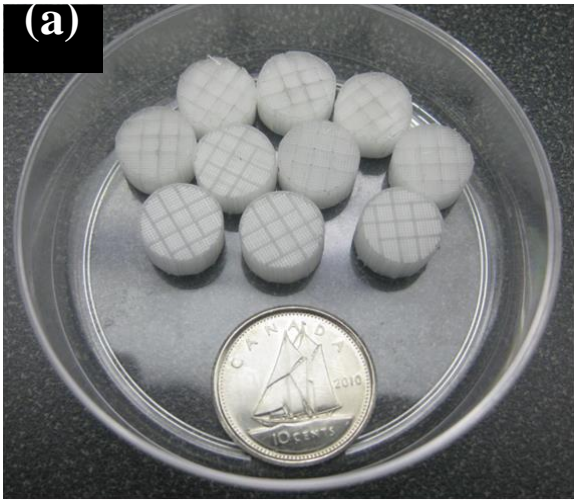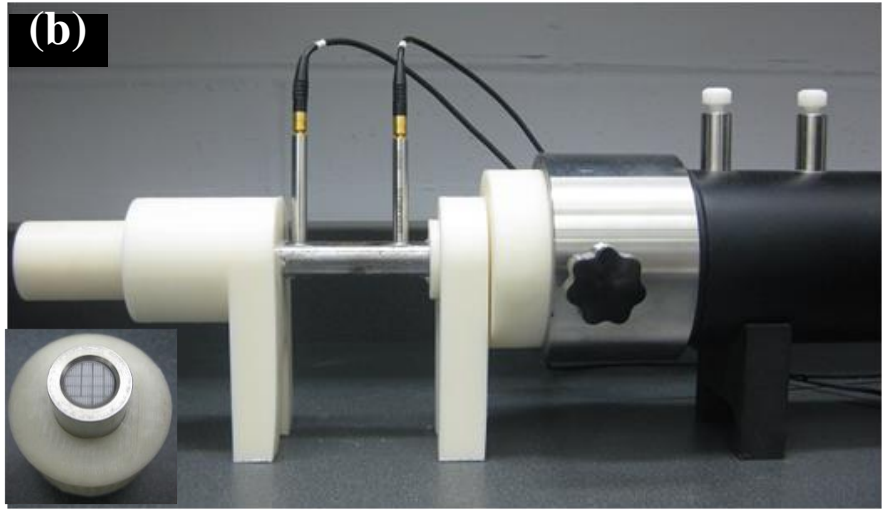
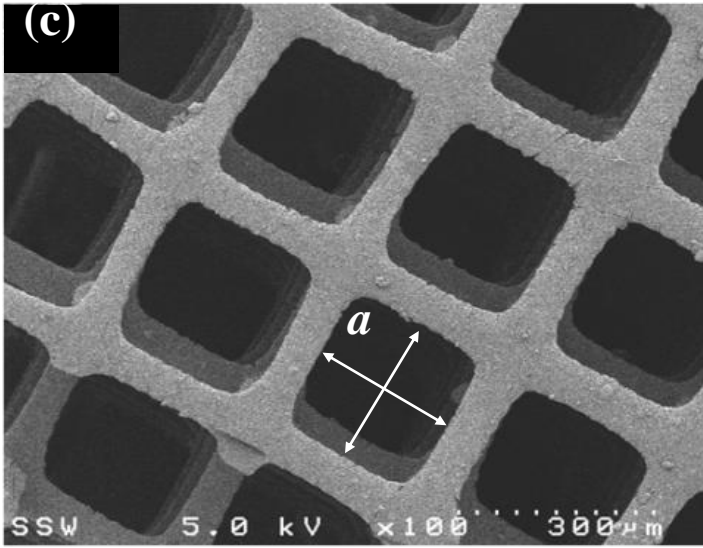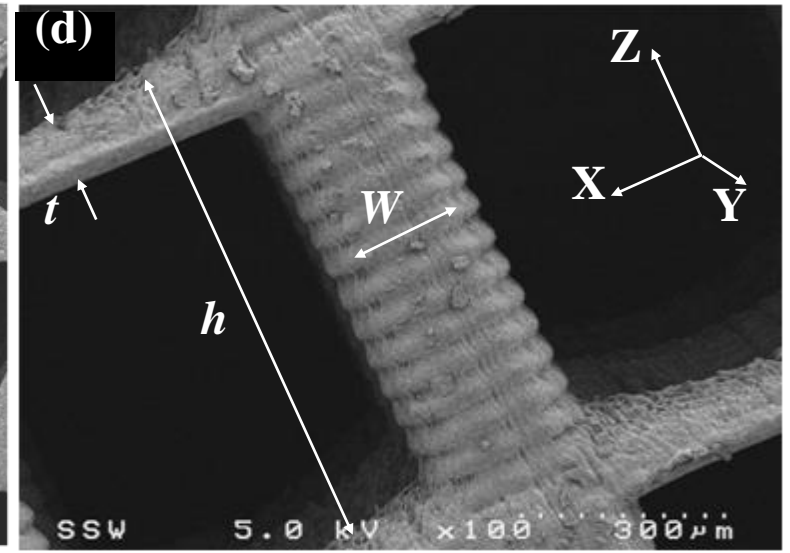

FIG.6

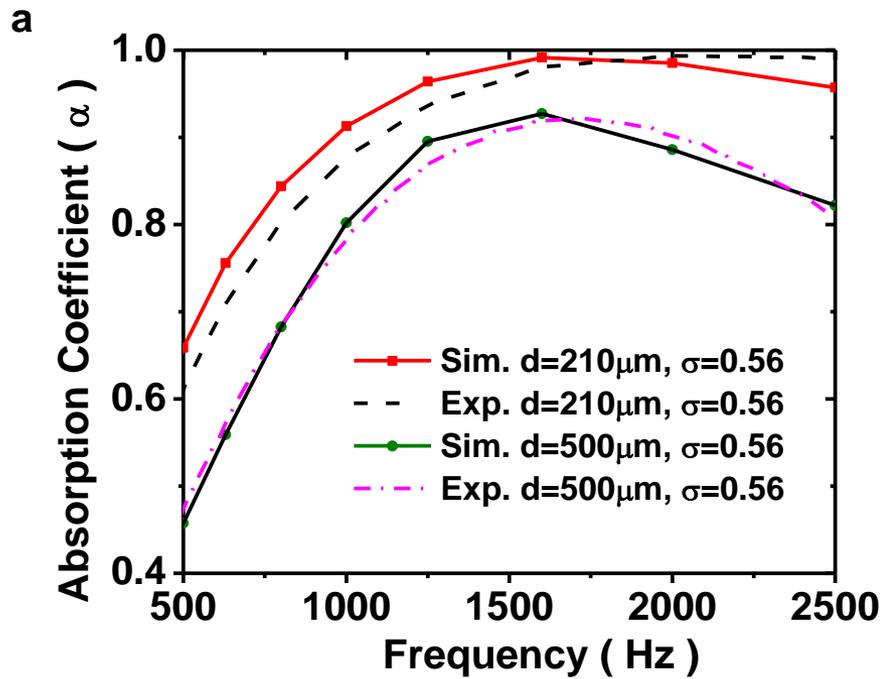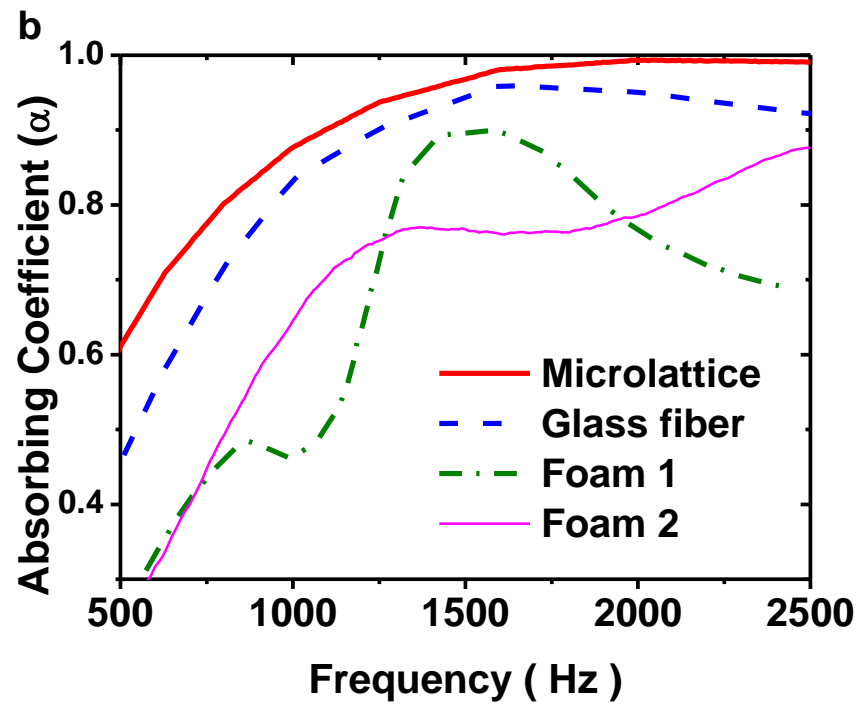

FIG.7